\documentclass[sn-aps,iicol]{sn-jnl}

\jyear{2021}%

\theoremstyle{thmstyleone}%
\theoremstyle{thmstyletwo}%

\theoremstyle{thmstylethree}%

\begin{document}

\title[Hypercyclic systems of measurements]{Hypercyclic systems of measurements and patterns of contextuality}


\author*[1]{\fnm{Víctor H.} \sur{Cervantes}}\email{victorhc@illinois.edu}

\author[2]{\fnm{Ehtibar N.} \sur{Dzhafarov}}


\affil[1]{\orgname{University of Illinois Urbana-Champaign},
\orgaddress{%
\city{Champaign},
\state{IL}, \country{USA}}}

\affil[2]{\orgname{Purdue University}, 
\orgaddress{%
 \city{West Lafayette},
 \state{IN}, \country{USA}}}



\abstract{%
We consider four measures of contextuality,
chosen for being based on the fundamental properties of the notion of contextuality,
and for being applicable to arbitrary
systems of measurements,
both without and with disturbance. 
We have previously shown that no two of them are functions of each other:
as systems of measurements change, either of them can change, while the other remains constant. 
This means that they measure different aspects of contextuality, 
and we proposed that rather than picking just one measure of contextuality in one specific sense, 
one could use all of them to characterize a contextual system by its pattern of contextuality. 
To study patterns of contextuality, however, one needs a systematic way of varying systems of measurements,
which requires their convenient parametrization. 
We have convenient parametrization within the class of cyclic systems that have played a dominant role in the foundations of quantum mechanics. 
However, they cannot be used to study patterns of contextuality, 
because within this class the four measures of contextuality have been shown to be proportional to each other. 
In this concept paper, we introduce hypercyclic systems of measurements. 
They generalize cyclic systems while preserving convenient parametrization. We show that within this class of systems, 
the same as for systems at large, no two of the measures of contextuality are functions of each other. 
This means that hypercyclic systems can be used to study patterns of contextuality.%
}




\maketitle


A system of random variables \(\mathcal{R}\) is a set of double-indexed random variables
\(R_{q}^{c}\), where \(q \in Q\) denotes their \emph{content},
that can be defined as the property the random variable measures,
and \(c \in C\) is their \emph{context}, encompassing the conditions
under which it is recorded. A system can be presented as 
\begin{equation}
\mathcal{R} = \{R_{q}^{c}: c \in C, q \in Q, q \prec c\}, \label{eq:sys}
\end{equation}
where \(q \prec c\) indicates that content \(q\) is measured in
context \(c\). The variables of the subset 
\begin{equation}
R^{c} = \{R_{q}^{c}: q \in Q, q \prec c\}
\end{equation}
are \emph{jointly distributed}, whereas any two random variables \(R_{q}^{c}, R_{q'}^{c'} \in \mathcal{R}\) with
\(c \neq c'\) are \emph{stochastically unrelated}, i.e., they possess no joint distribution. In particular, the variables in the subset
\begin{equation}
\mathcal{R}_{q} = \{R_{q}^{c}: c \in C, q \prec c\}
\end{equation}
are pairwise stochastically unrelated. The subset \(R^{c}\) is called the \emph{bunch} corresponding to context
\(c\), and the subset \(\mathcal{R}_{q}\) is referred to as the \emph{connection}
for content \(q\).

A \emph{consistently connected} system is a system for which
all variables within a connection have the same distribution.
In a \emph{strongly consistently connected} system,
for any \(c, c' \in C\), 

\begin{equation}
\{R_{q}^{c} : q \in Q, q \prec c, c'\} \overset{d}{=} \{R_{q}^{c'} : q \in Q, q \prec c, c'\}.
\end{equation}

A system (\ref{eq:sys}) is said to be \emph{noncontextual} \cite{Dzhafarov.2017.Contextuality}
if there is a set of identically labeled and \emph{jointly distributed} random variables 
\begin{equation}
S = \{S_{q}^{c}:c \in C, q \in Q, q \prec c\}, \label{eq:coup}
\end{equation}
satisfying the following two conditions.
(A) For any \(c \in C\), 
\begin{equation}
S^{c} = \{S_{q}^{c}: q \in Q, q \prec c\}\overset{d}{=} R^{c}. \label{eq:coup1}
\end{equation}
(B) For any \(q \in Q\), and any \(c, c' \in C\) such that \(q \prec c\) and \(q \prec c'\),
the probability of \(S_{q}^{c} = S_{q}^{c'}\) is maximal possible, given the individual distributions of the two variables.
If such a system (\ref{eq:coup}) does not exist, the system is \emph{contextual}.

\begin{figure*}
\caption{Contextuality measures of some hypercyclic systems of ranks 4 and 5}

\label{fig:Rank45}
\begin{center}
\includegraphics[width = 0.8\textwidth]{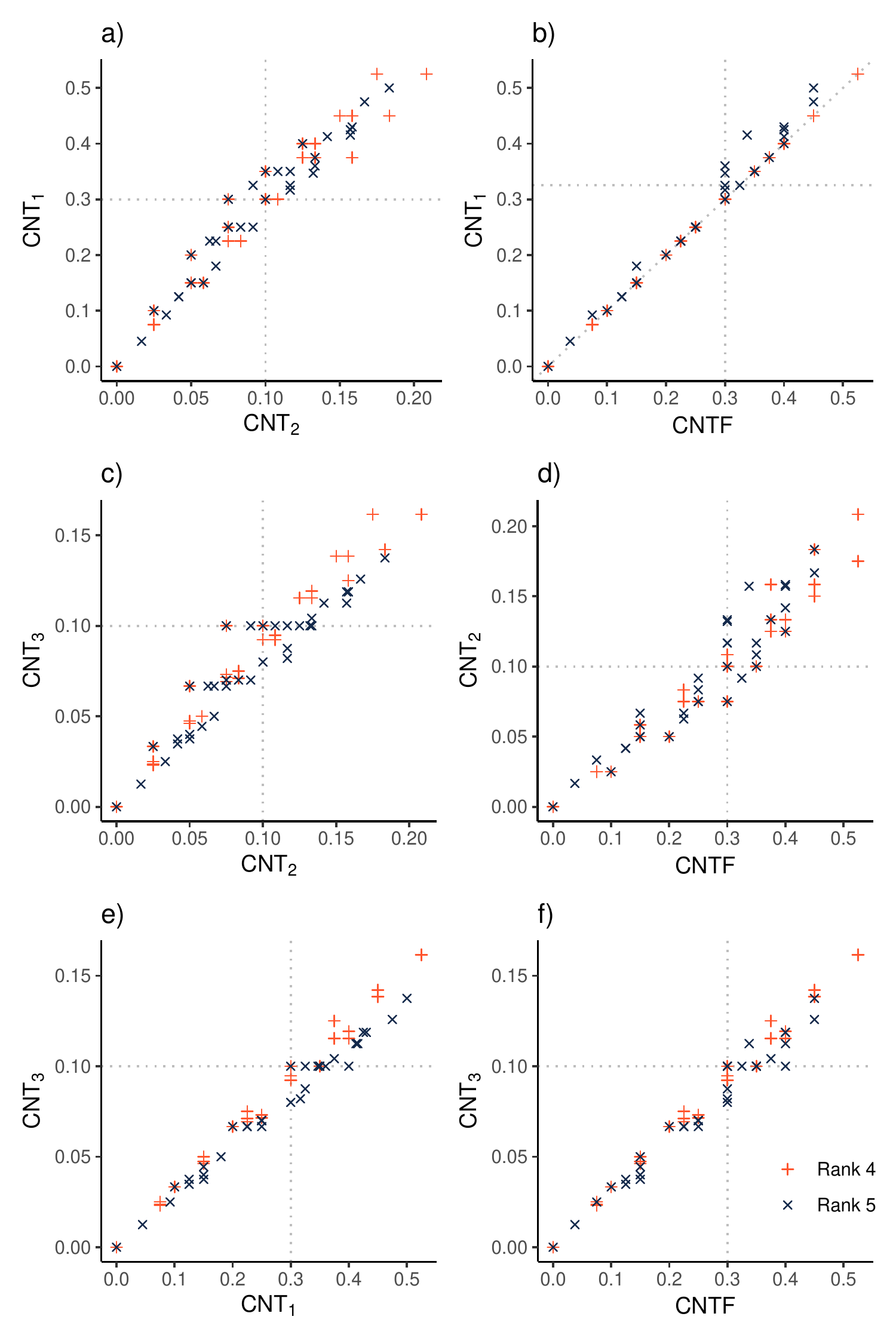}
\end{center}
{\footnotesize{
Note. Panel a) shows \(\text{CNT}_{1}\) vs \(\text{CNT}_{2}\)
for several hypercyclic systems of order 3 and ranks 4 and 5.
Panels b) through f) show, for the same systems, the other pairs of CNTs.
All variables in the systems used are dichotomous (0/1) and pairwise independent, with
\(\Pr(R_{q}^{c} = 1) = .5\).
Hence, for all systems the hierarchical measure is \(\text{CNT}_{2}^{3}\) and 
it coincides with \(\text{CNT}_{2}\).
The horizontal and vertical lines highlight examples for which
one of the measures remains constant while the other measure varies, proving that they are not related to each other by any function.
}}
\end{figure*}

We consider four principled and general measures of the degree of contextuality
(CNTs) proposed in the contextuality literature.
Each of them is constructed by relaxing one of the three basic constraints
defining noncontextual systems: that \(S\) is a probability distribution,
that it satisfies (A), and that it satisfies (B).

\(\text{CNT}_{1}\) measure
is computed by considering all distributions of \(S\) subject to (A),
quantifying their deviations from (B),
and taking the smallest such deviation for \(\text{CNT}_{1}\) \cite{Dzhafarov.2020.Contextuality}.
\(\text{CNT}_{2}\)
is defined by exchanging the places of (A) and (B)
in the previous sentence \cite{Dzhafarov.2020.Contextuality}.

\(\text{CNT}_{3}\) (proposed in Ref.~\cite{Abramsky.2011.sheaf-theoretic}
for strongly consistently connected systems,
and generalized in Ref.~\cite{Dzhafarov.2017.Contextuality})
is computed by replacing the distribution of \(S\) by a \emph{signed sigma-additive measure},
one that is allowed to be negative but equals \(1\) for the entire space of values of \(S\).
\(\text{CNT}_{3}\) equals \(T - 1\), where \(T\) is the smallest \emph{total variation}
among the signed measures subject to both (A) and (B).

One can also allow \(S\) to be a \emph{defective random variable},
with a nonnegative sigma-additive measure whose maximum value \(M\) can be less than \(1\).
This contextuality degree, called Contextual Fraction (\(\text{CNTF}\))
has been developed for strongly consistently connected systems in Ref.~\cite{Abramsky.2017.Contextual}.
It is the smallest value of \(1 - M\) among all such \(S\) subject to (B)
(with all maximal probabilities equal to \(1\)) and bounded from above by (A).
The latter means that the probabilities in the distribution of \(S^{c}\)
do not exceed the corresponding probabilities in the distribution of \(R^{c}\).
\(\text{CNTF}\) has been generalized to arbitrary systems in Ref.~\cite{Dzhafarov.2019.Contextuality-by-Default}.

For completeness, we should mention the hierarchical version of \(\text{CNT}_{2}\),
denoted \(\text{CNT}_{2}^{m}\),
which breaks down the contextuality of the system by the `levels' of the bunches
to find out at which level the incompatibility of (A) with (B) arises \cite{Cervantes.2020.Contextuality}.
The term `level', \(m\) in \(\text{CNT}_{2}^{m}\), refers to the number
of variables within each bunch
whose joint distribution is being considered.
For example, if the system
\[
\begin{array}{|c|c|c|c||c|}
\hline R_{1}^{1} & R_{2}^{1} & R_{3}^{1} & R_{4}^{1} & c = 1 \\
\hline R_{1}^{2} & R_{2}^{2} & R_{3}^{2} & R_{4}^{2} & \:\:2 \\
\hline R_{1}^{3} & R_{2}^{3} & R_{3}^{3} & R_{4}^{3} & \:\:3 \\
\hline R_{1}^{4} & R_{2}^{4} & R_{3}^{4} & R_{4}^{4} & \:\:4 \\
\hline
\hline q = 1     & 2         & 3         & 4         & \mathcal{R}_{4,4} \\
\hline
\end{array}
\]
is contextual at level 2, then the joint distributions of pairs of variables (2-marginals) in the system are already incompatible with (B); if such system is contextual at level 3, the 2-marginals are compatible with (B), but the 3-marginals of the bunches are not.

It has been documented in Ref.~\cite{Kujala.2019.Measures}
that generally, no two of the mentioned measures of contextuality
are functions of each other: as systems of measurements change,
either of them can change while the other remains constant.
This means that each of these CNTs 
generally depicts a unique aspect of contextuality.
For instance, different CNTs may be related to different geometric representations
of the same class of systems \cite{Kujala.2019.Measures}.
It seems, therefore, that rather than trying to find which of these CNTs is best,
one could use all of them as describing what can be called
\emph{pattern of contextuality}.
To systematically study such patterns, one needs a sufficiently rich but conveniently
parametrized class of systems of measurement.
We propose to use for this purpose a class of systems we dubbed \emph{hypercyclic}.

This class includes the well-known cyclic systems as a proper subset.
A cyclic system of random variables is a system for which
a suitable arrangement of the contents and contexts
allows one to present it as 
\begin{equation}
\mathcal{R}_{n} = \left\{ \left\{ R_{i}^{i}, R_{i \oplus 1}^{i} \right\} : i = 1, \ldots, n \right\}, \label{eq:cyclicSystem}
\end{equation}
where \(\oplus 1\)
denotes cyclic shift \(1 \mapsto 2, \ldots, n - 1 \mapsto n, n \mapsto 1\).
In (\ref{eq:cyclicSystem}), the variables \(\left\{ R_{i}^{i}, R_{i \oplus 1}^{i} \right\}\)
constitute the bunch corresponding to context \(c = i\).
It is clear that a cyclic system has the same number of contents and contexts. This number is called the \emph{rank} of the cyclic system.
The following
matrices depict three cyclic systems: of rank 2,
rank 3, and rank 4.
\[
\begin{array}{ccc}
\begin{array}{|c|c||c|}
\hline R_{1}^{1} & R_{2}^{1} & c = 1 \\
\hline R_{1}^{2} & R_{2}^{2} & c = 2 \\
\hline
\hline q = 1     & q = 2     & \mathcal{R}_{2} \\
\hline
\end{array}
&  &
\begin{array}{|c|c|c||c|}
\hline R_{1}^{1} & R_{2}^{1} &           & c = 1 \\
\hline           & R_{2}^{2} & R_{3}^{2} & \:\:2 \\
\hline R_{1}^{3} &           & R_{3}^{3} & \:\:3 \\
\hline
\hline q = 1     & 2         & 3         & \mathcal{R}_{3} \\
\hline
\end{array}
\end{array}
\]
\begin{equation}
\begin{array}{|c|c|c|c||c|}
\hline R_{1}^{1} & R_{2}^{1} &           &           & c = 1 \\
\hline           & R_{2}^{2} & R_{3}^{2} &           & \:\:2 \\
\hline           &           & R_{3}^{3} & R_{4}^{3} & \:\:3 \\
\hline R_{1}^{4} &           &           & R_{4}^{4} & \:\:4 \\
\hline
\hline q = 1     & 2         & 3         & 4         & \mathcal{R}_{4} \\
\hline
\end{array}
\end{equation}

Cyclic systems have played a very prominent role in the foundations of quantum physics,
especially in studies of contextuality and nonlocality
\cite{Araujo.2013.All,Kujala.2015.Necessary}.
However, in cyclic systems, due to their simplicity, all CNTs
mentioned above are related by simple proportionality
\cite{Dzhafarov.2020.Contextuality,Cervantes.2023.note,Camillo.2023.Measures}.
They are not, therefore, suitable for a systematic analysis of patterns of contextuality. 

A hypercyclic system of order \(k\) and rank \(n\), \(n \geq k\), is one
for which it is possible to enumerate its contents and contexts
to obtain the following representation: 
\begin{equation}
\mathcal{R}_{k, n} = \left\{ \left\{ R_{i}^{i}, R_{i \oplus 1}^{i}, \ldots, R_{i \oplus (k - 1)}^{i} \right\} : i = 1, \ldots, n \right\}, \label{eq:hypercyclicSystem}
\end{equation}
where \(\oplus j\) denotes \(j\) successive applications of cyclic shift \(\oplus 1\).
Similar to cyclic systems, a hypercyclic system has the same number of contents
and contexts. We continue to refer to this number as the rank of the system. 
The following
matrices depict three hypercyclic systems of order 3 and ranks 3, 4, and 5.

\[
\begin{array}{ccc}
\begin{array}{|c|c|c||c|}
\hline R_{1}^{1} & R_{2}^{1} & R_{3}^{1} & c = 1 \\
\hline R_{1}^{2} & R_{2}^{2} & R_{3}^{2} & \:\:2 \\
\hline R_{1}^{3} & R_{2}^{3} & R_{3}^{3} & \:\:3 \\
\hline
\hline q = 1     & 2         & 3         & \mathcal{R}_{3, 3} \\
\hline
\end{array}
&  &
\begin{array}{|c|c|c|c||c|}
\hline R_{1}^{1} & R_{2}^{1} & R_{3}^{1} &           & c = 1 \\
\hline           & R_{2}^{2} & R_{3}^{2} & R_{4}^{2} & \:\:2 \\
\hline R_{1}^{3} &           & R_{3}^{3} & R_{4}^{3} & \:\:3 \\
\hline R_{1}^{4} & R_{2}^{4} &           & R_{4}^{4} & \:\:4 \\
\hline
\hline q = 1     & 2         & 3         & 4         & \mathcal{R}_{3, 4} \\
\hline
\end{array}
\end{array}
\]
\begin{equation}
\begin{array}{|c|c|c|c|c||c|}
\hline R_{1}^{1} & R_{2}^{1} & R_{3}^{1} &           &           & c = 1 \\
\hline           & R_{2}^{2} & R_{3}^{2} & R_{4}^{2} &           & \:\:2 \\
\hline           &           & R_{3}^{3} & R_{4}^{3} & R_{5}^{3} & \:\:3 \\
\hline R_{1}^{4} &           &           & R_{4}^{4} & R_{5}^{4} & \:\:4 \\
\hline R_{1}^{5} & R_{2}^{5} &           &           & R_{5}^{5} & \:\:5 \\
\hline
\hline q = 1     & 2         & 3         & 4         & 5         & \mathcal{R}_{3, 5} \\
\hline
\end{array}
\end{equation}


Our numerical exploration of hypercyclic systems
shows that our CNTs
are not generally related to each other by any function.
Figure~\ref{fig:Rank45} illustrates this for
all six pairs of the CNTs computed for hypercyclic
systems of order 3 and ranks 4 and 5.
Based on this finding and the fact that hypercyclic systems can be systematically varied
due to their parametrization by order and rank,
one can hope that this class of systems could be helpful in studying patterns of contextuality.
In this concept paper, we cannot assert more.

\section*{Acknowledgements}

This research was partially supported by the Foundational Questions Institute grant FQXi-MGA-2201.

\section*{Data availability}

No data are associated with the manuscript.


\end{document}